\documentstyle[aps,12pt]{revtex}

\begin{document}
\vspace{2cm}
\title{Enhancement of Quantum Tunneling for Excited States in Ferromagnetic Particles}

\vspace{3,5cm}
\author{J.-Q. Liang$^{1,2,3}$, Y.-B. Zhang$^{3,2}$, H. J. W. M\"{u}ller-Kirsten$^1$,
Jian-Ge Zhou$^{1,4}$, F. Zimmerschied$^1$, F.-C. Pu$^{5,3}$}
\address{$^1$Department of Physics, University of Kaiserslautern, 67653\\
Kaiserslautern, Germany\\
$^2$Institute of Theoretical Physics, Shanxi University, Taiyuan, Shanxi
030006, P. R. China\\
$^3$Institute of Physics and Center for Condensed Matter Physics, Chinese
Academy of Sciences, Beijing 100080, P. R. China\\
$^4$Department of Physics, Osaka University, Toyonaka, Osaka 560, Japan\\
$^5$Department of Physics, Guangzhou Normal College, Guangzhou 510400, P. R.
China}
\maketitle

\vspace{3.5cm}
\begin{abstract}
A formula suitable for a quantitative evaluation of the tunneling
effect in a ferromagnetic particle is derived with the help of the instanton
method. The tunneling between $n$-th degenerate states of neighboring wells
is dominated by a periodic pseudoparticle configuration.
The low-lying level-splitting previously obtained with the LSZ method
in field theory in which the tunneling is viewed as the transition
of $n$ bosons induced by the usual(vacuum) instanton is recovered.The
observation made with our new result
is that the tunneling effect increases at excited states. The
results should be useful in analyzing results of experimental tests of
macroscopic quantum coherence in ferromagnetic particles.
\end{abstract}

\newpage
\section{Introduction}

Macroscopic quantum effects in magnetic systems are of considerable interest
both theoretically and experimentally\cite{1}. In the context of these
investigations the usual terminology is that macroscopic quantum coherence%
\cite{2} (MQC) refers to the resonance between neighboring degenerate wells.
Some years ago it was reported that MQC was observed for antiferromagnetic
particles \cite{3} in resonance experiments. However, the interpretation is
controversial\cite{2,3,4}. Physically the result of an earlier resonance
experiment on ferromagnetic particles is not clear since fundamental
discrepancies remain between the experimental data and theoretical
expectations on the basis of magnetic quantum tunneling \cite{5,6}. Apart
from some other reasons\cite{5,6}, which hinder the acceptance of the
observation as definite proof of MQC, there is an essential difficulty
related to the existing theory of quantum tunneling itself in the absence of
an external magnetic field. The difficulty was pointed out in ref.\cite{2}:
The argument of the WKB exponential of the tunneling for a ferromagnetic
particle is $2\sqrt{\lambda }s$ with $\lambda =\frac{K_2}{K_1}$, $K_1$ and $%
K_2$ being the hard and medium axis energies. Since $s\approx 500$ to $5000,$
unless $\frac{K_2}{K_1}\ll 10^{-4}$, the tunneling frequency $\omega _c$ is
expected to be unobservably small.

It seems that the unobservably small effect of tunneling at the vacuum level is
a common phenomenon in various problems, as for instance,
in the case of baryon- and lepton-number
violation at high energy and in the case of pair production of black holes in quantum
gravity. The one loop correction which results in a prefactor of the WKB
leading order exponential does not enhance the tunneling significantly in this case. It
is a natural and interesting question to ask whether the tunneling effect is
enhanced by considering tunneling at the level of excited states. However, the
instanton method is suitable only for the evaluation of the tunneling effect
at the vacuum level, since the usual(vacuum) instantons satisfy the vacuum
boundary conditions. Motivated by the study of baryon- and lepton-number
violation at high energy, recently new types of pseudoparticle
configurations were found\cite{7,8,9,10,11} which satisfy periodic boundary
conditions and are called periodic instantons\cite{8,9} or
nonvacuum instantons\cite{10,11}.
These periodic instantons have, for instance,been used to evaluate quantum
tunneling at high energy\cite{12}. There it was confirmed that in the low energy
region the tunneling effect indeed increases exponentially with energy\cite
{12}. This finding can be expected to have its correspondence in the
theoretical analysis of MQC. In the present paper we therefore adopt the
periodic instanton method in order to calculate the tunneling amplitude
between asymptotically degenerate excited states.

We derive a compact formula for the level-splitting induced by tunneling
which is valid for the entire region of energy. The results of the
application of a method \cite{13}
previously developed for the calculation of tunneling effects
at the level of excited states and based on the LSZ (Lehmann, Symanzik and
Zimmermann) procedure of field theory \cite{14} are recovered
in the low energy region \cite{15}. 
In particular our formula agrees exactly with the level splitting of the
ground state obtained by means of the usual instanton method\cite{16,17}. It is
remarkable that the tunneling effect enhances significantly if tunneling at
the level of an asymptotically degenerate excited state is considered.
We have shown elsewhere that in certain restricted parameter domains
the leading contributions of the
effect can also be obtained much more easily with Schr\"odinger
quantum mechanics\cite{19}, even in the presence of an applied magnetic field
\cite{20}.

\section{The effective Lagrangian with the periodic potential, and the
energy spectrum formula}

We begin with the following operator Hamiltonian of the ferromagnetic
particle which has been the starting point of numerous investigations
\begin{equation}
\hat{H}=K_1\hat{s}_z^2+K_2\hat{s}_y^2  \label{1}
\end{equation}
and describes \cite{16,17} XOY easy plane anisotropy and an easy axis along
the x direction with $K_1>K_2>0$. In eq.(\ref{1}) $\hat{s}_i,$ $i=x,y,z$,
are spin operators obeying the usual commutation relation $[\hat{s}_i,\hat{s}%
_i]=i\epsilon _{ijk}\hat{s}_k$ (using natural units throughout,i.e. $\hbar
=c=1$). Starting from the coherent state representation \cite{21} of the
time evolution operator with Hamiltonian given by eq.(\ref{1}) and with the
help of the coherent state path--integral we obtain 
\begin{equation}
\langle {\bf n}_f|e^{-2i\hat{H}T}|{\bf n}_i\rangle =e^{-i(\phi _f-\phi _i)s}%
{\cal K}(\phi _f,t_f;\phi _i,t_i)  \label{2}
\end{equation}
where 
\begin{equation}
{\cal K}(\phi _f,t_f;\phi _i,t_i)=\int {\cal D}\phi {\cal D}%
pe^{i\int_{t_i}^{t_f}{\cal L}(\phi ,p)dt}  \label{3}
\end{equation}
is the path integral in phase space with canonical variables $\phi $ and $%
p\equiv s\cos \theta $\cite{21}. Also 
\begin{equation}
{\cal L}=\dot{\phi}p-H(\phi ,p)  \label{4}
\end{equation}
is the phase space (or first order) Lagrangian. The Hamiltonian 
\begin{equation}
H=\frac{p^2}{2m(\phi )}+V(\phi )  \label{5}
\end{equation}
has position dependent mass $m(\phi )$ and potential 
\begin{equation}
m(\phi )=\frac 1{2K_1(1-\lambda \sin ^2\phi )},\quad V(\phi )=K_2s(s+1)\sin
^2\phi  \label{6}
\end{equation}
respectively where $\lambda \equiv \frac{K_2}{K_1}$. The kets $|{\bf n}_i>$
and $|{\bf n}_f>$ denote the initial and final spin--coherent states and $%
t_f-t_i\equiv 2T$ denotes the difference of final and initial times. Here $%
\vec{s}=s(\sin \theta \cos \phi ,\sin \theta \sin \phi ,\cos \theta )$ is
visualized as a classical spin vector with spin number $s$, polar angle $%
\theta $ and azimuthal angle $\phi $. In the above derivation, the large
spin limit $s\gg 1$ has been used since giant spins with spin quantum number 
$s\gg 1$ are believed to describe ferromagnetic grains. A novel
feature of the transition amplitude given by eq.(\ref{2}) is the phase
factor $e^{-i(\phi _f-\phi _i)s}$ which can be put into the Lagrangian, i.e.
the expression $\phi _f-\phi _i=\int_{t_i}^{t_f}\dot{\phi}dt$, and
identified as a Wess--Zumino term \cite{22}. Integrating out the momentum in
the path integral eq.(\ref{3}), we obtain the usual Feynman propagator in
configuration space, i.e. 
\begin{equation}
{\cal K}(\phi _f,t_f;\phi _i,t_i)=\int \tilde{{\cal D}}\phi
e^{i\int_{t_i}^{t_f}{\cal L}(\phi ,\dot{\phi})dt}  \label{7}
\end{equation}
where $\tilde{{\cal D}}\phi $ is the measure-modified functional
differential resulting from the $\phi $-dependent mass, i.e. 
\[
\tilde{{\cal D}}\phi =\prod_k^{M-1}\sqrt{\frac{m(\phi _k)}{2\pi i\epsilon }}%
d\phi _k 
\]
with the second order Lagrangian 
\begin{equation}
{\cal L}=\frac 12m(\phi )\ {\dot{\phi}}^2-V(\phi )  \label{8}
\end{equation}
which is more convenient for the instanton method used in the following. The
potential $V(\phi )$ is periodic with period $\pi $ (Fig. 1) and there are
two minima in the entire region $2\pi $. We may look at this periodic
potential as a superlattice with lattice constant $\pi $ and total length $%
2\pi $, and we can derive the energy spectrum in the tight-binding
approximation. The translational symmetry is ensured by the possibility of
successive $2\pi $ extensions.

It was shown in a previous paper\cite{23} that
if $\epsilon_m$ are the degenerate eigenvalues of the system
with infinitely high barriers, the energy spectrum is given
by 
\begin{equation}
E_m=\epsilon _m-2\triangle \epsilon _m\cos (s+\xi )\pi  \label{9}
\end{equation}
where the expression $\triangle \epsilon _m$ is defined by 
\begin{equation}
\triangle \epsilon _m=-\int u_m^{*}(\phi ,\Phi _n)\hat{H}u_m(\phi ,\Phi
_{n+1})d\phi  \label{10}
\end{equation}
which is the usual overlap integral or
$2\triangle \epsilon_m$ simply the level shift due to
tunneling through any one of the barriers. Here $u_m(\phi -\Phi _n)$ denotes
the eigenfunction of the harmonic oscillator-approximated Hamiltonian $\hat{H%
}_0$ in the $n$-th well, i.e. 
\[
H_0=\frac{p^2}{2m_0}+\frac 12m_0{\omega _0}^2(\phi -\Phi _n)^2 
\]
with $m_0=\frac 1{2K_1}$ and ${\omega _0}^2=4K_1K_2s(s+1).$ $\xi $ is an integer
and here can assume only either of the two values ``0'' and ``1''.
For half integer spin $s$
the spectrum eq. (\ref{9}) is quenched to a single degenerate level with
degeneracy two. The quenching is seen to be a consequence of Kramer's
theorem which says that for half integer spin $s$ the degeneracy cannot be
removed in the presence of the crystal field\cite{23}.

\section{Level splitting of ground state derived with the usual instanton
method}

Above we recalled the low-lying energy spectrum of a giant spin particle in
the large spin limit as obtained in\cite{23}. The energy spectrum eq.(%
\ref{9}) leads to the level splitting given by the absolute difference for$%
\xi =0$ and $1$, namely
\begin{equation}
\triangle E_m=2\triangle \epsilon _m\left| \cos (s+1)\pi -\cos s\pi \right|
=\{_{0\qquad \qquad \text{for half integer spin }s}^{4\triangle \epsilon
_m\qquad \text{for integer spin }s}  \label{11}
\end{equation}
The only parameter left unknown is the overlap integral or the level shift $%
2\triangle \epsilon _m$ which can be evaluated with the help of the instanton
method. Instantons in field theory of 1+0 dimensions are viewed as
pseudoparticles with trajectories existing in barriers, and are therefore
responsible for tunneling. Since instanton trajectories interpolate between
degenerate vacua (see Fig. 1 (a)) and satisfy vacuum boundary conditions, the
instanton method is only suitable for the calculation of tunneling splitting
$4\triangle\epsilon_0$ between
neighboring vacua. In the following we first consider tunneling at the
vacuum level (i.e. $m=0$), which leads to the level shift of the ground state energy,
i.e. $2\triangle \epsilon _0$ of eq.(\ref{10}). Passing to imaginary time by
Wick rotation $\tau =it,\beta =iT,$ the amplitude for tunneling from the
initial well, say that with $n=0$ (and $\phi _i=0$), to the neighboring well
with $n=1$ (and $\phi _f=\pi $) and considering large $\beta $, the
amplitude for the transition between the corresponding coherent states
can be shown to be 
\begin{eqnarray}
<{\bf n}(\pi )|e^{-2\beta \hat{H}}|{\bf n}(0)> &=&<{\bf n}(\pi )|0,\Phi
_1><0,\Phi _0|{\bf n}(0)>e^{-2\beta \epsilon _0}\sinh (2\beta \triangle
\epsilon_0 )  \nonumber \\
&=&e^{-i\pi s}{\cal K}_E(\phi _f=\pi ,\beta ;\phi _i=0,-\beta )  \label{12}
\end{eqnarray}
where 
\[
{\cal K}_E=\int {\cal D}{\phi }e^{-S_E}
\]
is the Euclidean propagator with Euclidean action defined by 
\begin{equation}
S_E=\int_{-\beta }^\beta {\cal L}_Ed\tau ,\;\ {\cal L}_E=\frac 12m(\phi )%
\dot{\phi}^2+V(\phi )  \label{13}
\end{equation}
A relation similar to eq.(\ref{12}) applies for a transition between
asymptotically degenerate states ``$m$'' with $\epsilon_o$ replaced by
$\epsilon_m$.  Such transitions will 
be considered in the next
section, and the splitting will similarly be read off from eq.(\ref{12}).
From now on $\dot{\phi}=\frac{d\phi }{d\tau }$ denotes the imaginary time
derivative.

In the following the Euclidean propagator ${\cal K}_E$ is evaluated with the
instanton method. After evaluation we compare the result with eq.(\ref{14})
to find the level shift $2\triangle \epsilon _0$. The instanton configuration
which minimizes the Euclidean action $S_E$ is 
\begin{equation}
\phi _c=\arcsin [\cosh ^2\omega _0(\tau -\tau _0)-\lambda \sinh ^2\omega
_0(\tau -\tau _0)]^{-\frac 12}  \label{14}
\end{equation}
with position $\tau _0$. The instanton trajectory is shown in Fig. 1 (a)
with $\tau _0=0.$ The Euclidean action evaluated for the instanton
trajectory eq.(\ref{14}), sometimes called the instanton mass, is 
\begin{equation}
S_c=\int_{-\infty }^\infty m(\phi _c)\dot{\phi}_c^2d\tau =\sqrt{s(s+1)}\ln 
\frac{1+\sqrt{\lambda }}{1-\sqrt{\lambda }}\approx (s+\frac 12)\ln \frac{1+%
\sqrt{\lambda }}{1-\sqrt{\lambda }}  \label{15}
\end{equation}
in agreement with refs.\cite{16,17} and with refs.
\cite{19,20} in the case of small values of $\lambda$.
To our understanding it is argued in refs.\cite{16,17} that
except in $e^{-S_c}$ one can replace $s(s+1)$ by $s^2$,
whereas in the exponential factor $\sqrt{s(s+1)}$ has to be
approximated by $s+\frac{1}{2}$, even for large $s$.
The functional integral ${\cal K}_E$
can be evaluated with the stationary phase method by expanding $\phi $ about
the instanton trajectory $\phi _c$ such that $\phi =\phi _c+\eta $, where $%
\eta $ is the small fluctuation with boundary conditions $\eta (\beta )=\eta
(-\beta )=0$. Up to the one-loop approximation we have 
\begin{equation}
{\cal K}_E=e^{-S_c}I  \label{16}
\end{equation}
where 
\begin{equation}
I=\int_{\eta (-\beta )=0}^{\eta (\beta )=0}{\cal D}\eta e^{-\delta S_E}
\label{17}
\end{equation}
is the fluctuation functional integral with the fluctuation action 
\begin{equation}
\delta S_E=\int_{-\beta }^\beta \eta \hat{M}\eta d\tau   \label{18}
\end{equation}
where 
\begin{equation}
\hat{M}=-\frac 12\frac d{d\tau }m(\phi _c)\frac d{d\tau }+\tilde{V}(\phi _c)
\label{19}
\end{equation}
with 
\begin{equation}
\tilde{V}(\phi _c)=\frac 12[-m^{\prime }(\phi _c)\ddot{\phi}_c-\frac 12%
m^{\prime \prime }(\phi _c)\dot{\phi}_c^2+V^{\prime \prime }(\phi _c)]
\label{20}
\end{equation}
Here $\hat{M}(\phi _c)$ is the operator of the second variation of the
action and $m^{\prime }(\phi _c)={\frac{\partial m(\phi )}{\partial \phi }}%
|_{\phi =\phi _c}$. As in the usual method of evaluating the fluctuation
integral $I$, we expand the fluctuation variable $\eta $ in terms of the
eigenmodes of $\hat{M}$ and set $\eta =\Sigma _nC_n\psi _n$, where $\psi _n$
denotes the $n$-th eigenfunction of $\hat{M}$, and express the result of the
integration as an inverse square root of the determinant of $\hat{M}$. In
view of the time translation symmetry of the equation of motion, the
functional integral ${\cal K}_E$ is not well defined when expanded about the
classical solutions. The translational symmetry results in zero eigenmodes
of the second variation operator $\hat{M}$ of the action (which in the
present case, of course, has only one). This problem can be cured by the
Faddeev--Popov procedure \cite{24} or in a more systematic way with the help
of the BRST transformation \cite{25}. Following the procedure of refs.\cite
{23,25} the one instanton contribution to the propagator in the one-loop
approximation is calculated to be 
\begin{equation}
{\cal K}_E^{(1)}=2\beta \frac 4\pi (1-\lambda )^{-\frac 12}s^2K_2e^{-\omega
_0\beta }e^{-S_c}  \label{21}
\end{equation}
To obtain the desired result proportional to $\sinh (2\beta \triangle
\epsilon _0)$ in eq.(\ref{12}), the contributions of the infinite number of
instanton and anti-instanton pairs to the one instanton contribution have to
be taken into account (the trajectory of one instanton plus a pair is shown
in Fig. 1 (b)). Interactions among instantons and anti-instantons are
neglected in the dilute instanton--gas approximation. The contribution of
one instanton plus $n$ such pairs to the propagator is obtained with the
help of the group property of the Feynman path--integral and is found to be 
\begin{equation}
{\cal K}_E^{(2n+1)}=\frac{(2\beta )^{2n+1}}{(2n+1)!}(\frac s\pi )^{\frac 12%
}\lambda ^{\frac 14}[2^2\{\frac{K_1K_2}{(1-\lambda )\pi }\}^{\frac 12%
}\lambda ^{\frac 14}s^{\frac 32}]^{2n+1}e^{-(2n+1)S_c}e^{-\omega _0\beta }
\label{22}
\end{equation}
Summing over all contributions ${\cal K}_E^{(2n+1)}$, the final result of
the propagator is found to be 
\begin{equation}
{\cal K}_E=\lambda ^{\frac 14}(\frac s\pi )^{\frac 12}e^{-\beta \omega
_0}\sinh [2\beta .2^2\{\frac{K_1K_2}{(1-\lambda )\pi }\}^{\frac 12}\lambda ^{%
\frac 14}s^{\frac 32}e^{-S_c}]  \label{23}
\end{equation}
Comparing with eq.(\ref{12}) the level shift is seen to be $2\triangle
\epsilon _0$ (note that this is the shift a single level)  with 
\begin{equation}
\triangle \epsilon _0=2^2\{\frac{K_1K_2}{(1-\lambda )\pi }\}^{\frac 12%
}\lambda ^{\frac 14}s^{\frac 32}e^{-(s+\frac 12)\ln \frac{1+\sqrt{\lambda }}{%
1-\sqrt{\lambda }}}.  \label{24}
\end{equation}
Expanding $\triangle \epsilon _0$ in the region of small values of $\lambda $%
, this is 
\begin{equation}
\triangle \epsilon _0=2^2\{\frac{K_1K_2}\pi \}^{\frac 12}\lambda ^{\frac 14%
}s^{\frac 32}e^{-2s\sqrt{\lambda }}  \label{25}
\end{equation}
the level splitting being $4\triangle\epsilon_0$.
Since $s$ is a large number ($500$ to $5000$ as cited in the literature \cite
{3}), the level shift is suppressed in leading order by the factor 
\begin{equation}
e^{-S_c}\simeq e^{-2s\sqrt{\lambda }}  \label{26}
\end{equation}
For $\triangle \epsilon _0$ to be observable, plausible values of $\lambda $
are of order $10^{-5}.$ Such small values of $\lambda $ may not be
realizable in nature. The quantum correction proportional to $e^{\frac 32\ln
s}$ does not increase the tunneling probability significantly.

It may be interesting to compare our level splitting of the ground state $\Delta
E_0$ with that in refs. \cite{16,17}. The Hamiltonian of eq.(1) in refs. \cite{16}
and \cite{17} equals that of our model if one sets the external field equal
to zero ($h=0$%
) and makes the replacements $A=K_2,B=K_1-K_2$.
By simple algebra one can find
that the level splitting eq.(16) in ref.\cite{17} (or eq.(9a) in ref. \cite
{16}) coincides exactly with our $\Delta E_0=4\triangle \epsilon _0$ which
also simplifies to the results of refs. \cite{19,20,23}in
the appropriate limits.
We can now proceed to consider nonvacuum or periodic instantons which are
the pseudoparticles interpolating between asymptotically degenerate
excited states, and which reproduce the above result for vacuum instantons
in the vacuum limit.

\section{Quantum tunneling for excited states and the generalized formula for
the level splitting}

The periodic instanton method\cite{8,9,10,11,12} has become a
powerful tool for the evaluation of quantum tunneling amplitudes
over the entire region of
energy. The model at hand can be looked at as one for
tunneling at the level of excited states of
a sine--Gordon--type potential with a position-dependent mass,which has not been
reported previously in the literature. The periodic instanton configuration which
minimizes the Euclidean action eq. (\ref{13}) is seen to satisfy the
equation of motion 
\begin{equation}
\frac{m(\phi _p)}2\left( \frac{d\phi _p}{d\tau }\right) ^2-V(\phi _p)=-E_{cl}
\label{27}
\end{equation}
where $E_{cl}>0$, which is a constant of integration, may be viewed as the
classical energy of the pseudoparticle configuration. Through the usual
procedure of derivation of the periodic instanton solution, after a
laborious but straightforward integration of eq. (\ref{27}), we obtain the
periodic instanton configuration\cite{27} 
\begin{equation}
\phi _p=\arcsin \left[ \frac{1-k^2%
\mathop{\rm sn}
^2\left( \omega \tau |k\right) }{1-\lambda k^2%
\mathop{\rm sn}
^2\left( \omega \tau |k\right) }\right] ^{\frac 12}  \label{28}
\end{equation}
where $%
\mathop{\rm sn}
\left( \omega \tau |k\right) $ denotes a Jacobian elliptic function of
modulus $k$, 
\begin{equation}
k=\sqrt{\frac{n_1^2-1}{n_1^2-\lambda }}  \label{29}
\end{equation}
with 
\begin{equation}
n_1^2=\frac{K_2s(s+1)}{E_{cl}},\qquad \omega =\omega _0\sqrt{1-\frac{\lambda
E_{cl}}{K_2s(s+1)}}  \label{30}
\end{equation}

One can check with $%
\mathop{\rm sn}
\left(u|1\right)=\tanh u$
that for $k^2 = 1$ (corresponding to $E_{cl}\rightarrow 0$)
the configuration (\ref{28}) reduces to
the instanton of eq.(\ref{14}).  The Jacobian elliptic function $%
\mathop{\rm sn}
$ has real periods $4n{\cal K}(k)$, $n$ being an integer and
${\cal K}(k)$ the quarter period given by the usual
elliptic integral of the first kind.  The parameter values
(\ref{29}), (\ref{30}) ensure the periodicity
of $\phi_p=0$ at $\tau = -2\beta, +2\beta$ with a
crossover from negative to positive values at $\tau = 0$ and
$\phi = \frac{\pi}{2}mod 2\pi$.
Thus the one--way transition from a turning point $a_1$ to
the other turning point $a_2$ is
mediated by the instanton--like part or one half of the periodic
instanton extending from $\tau = -\beta$ to $\tau = +\beta$
(the periodic instanton itself returning to its original position after
time $4\beta$). 
The trajectory of the periodic instanton eq.(\ref{28}) is shown
schematically in
Fig. 1(c) where the trajectory is shifted by an amount $2\pi $. The
instanton--like part starts
at time $-\beta$ from turning point $\phi = a_1=\arcsin \sqrt{\frac{E_{cl}}{K_2s(s+1)}}$
and reaches the other turning point $\phi = a_2=\pi -\arcsin \sqrt{\frac{E_{cl}}{%
K_2s(s+1)}}$ at time $\beta$. The Euclidean action of the periodic instanton
configuration eq.(\ref{28}) over the domain from $\tau = -\beta$
to $\tau = +\beta$ can be found to be (cf. refs.\cite{9,10}) 
\begin{equation}
S_p=\int_{-\beta}^{\beta}\left[ m(\phi _p)\dot{\phi}_p^2+E_{cl}\right] d\tau
=W+2E_{cl}\beta  \label{31}
\end{equation}
where 
\begin{equation}
W=\frac \omega {\lambda K_1}\left[{\cal K}(k)-(1-\lambda k^2)\Pi (\lambda
k^2,k)\right]   \label{32}
\end{equation}
and $\Pi (\lambda k^2, k)$ is the complete elliptic integral of the
third kind.

It is now necessary to calculate the amplitude $A^{(1)}$ for the transition mediated
by one pseudoparticle configuration -- in the present case
the instanton--part of the periodic instanton -- and then to
sum over this contribution with other contributions $A^{(2n+1)}$ obtained by
adding an arbitrary number ($n=1,2,3,...$) of noninteracting complete
periodic instantons
(corresponding to addition of instanton--anti-instanton pairs
in the dilute gas approximation
of the instanton procedure) so that in the path integral representation
of the amplitude all possible paths between $-\beta$ and $+\beta$
are taken into account.  As is well known, this results effectively in an
exponentiation of the first contribution (see e.g. ref.\cite{11}). 
A calculation very similar to this can be found in
reference \cite{10} (cf. there eqs. (4.19), (4.24), (4.25)).  We
can therefore simply transcribe the result into the
present context.  The total amplitude corresponding
to eq.(\ref{12}) is then
$$
A = \sum^{\infty}_{m=0} A^{(2m+1)} = e^{-2E_{cl}\beta}\sinh\left\{\frac{2\beta\omega}{4
{\cal K}(k')} e^{-W}\right\}
$$
Comparison with eq. (\ref{12}) as noted earlier then yields the
compact formula\cite{8,9,10} 
\begin{equation}
\Delta E_{cl}=\frac \omega {4{\cal K}(k^{\prime })}e^{-W}  \label{33}
\end{equation}
$k^{\prime }$ being the complementary modulus of $k$, i.e. 
$%
k^{\prime }=\sqrt{1-k^2}.$ We emphasize that the above formula eq. (\ref{33}%
) obtained with the periodic instanton is valid for the entire region of
energy $E_{cl}$. The validity of the level splitting of excited states 
considered previously in the
literature\cite{28} was restricted to the quasi uniaxial limit, $%
1-\lambda \ll 1.$ Our approach,  however, is valid over the whole range of $%
\lambda .$

We now consider the low energy limit where $E_{cl}$ is much less than the
sphaleron energy (or
barrier height $K_2s(s+1)$ of the potential),
i.e. $k\rightarrow 1,k^{\prime }\rightarrow 0.$
Expanding $W$ as power series of $k^{\prime }$ up to quadratic order as in
ref. \cite
{9} and making use of the oscillator approximation of the periodic potential
around one of the minima with the quantization
replacement $E_{cl}\rightarrow\epsilon_m =\left( m+\frac 12\right) \omega_0 $%
(as in refs. \cite{8,10}), we then have 
\begin{equation}
W=(s+\frac 12)\ln \frac{1+\sqrt{\lambda }}{1-\sqrt{\lambda }}+\left( m+\frac 
12\right) \ln \left[ \frac{1-\lambda }{8\sqrt{\lambda }s}\left( m+\frac 12%
\right) \right] -\left( m+\frac 12\right)   \label{34}
\end{equation}
The level shift $2\Delta \epsilon _m$ is finally given with 
\begin{equation}
\Delta \epsilon _m=\frac 1{m!}2^{3m+2}\frac{\lambda ^{\frac 12(m+\frac 12%
)}s^{m+\frac 32}}{\left( 1-\lambda \right) ^m}\left\{ \frac{K_1K_2}{\pi
\left( 1-\lambda \right) }\right\} ^{\frac 12}\exp \left[ -(s+\frac 12)\ln 
\frac{1+\sqrt{\lambda }}{1-\sqrt{\lambda }}\right]   \label{35}
\end{equation}
which reduces to the ground state expression eq.(\ref{24}) for $m=0$. Eq. (%
\ref{35}) provides the linkage from the splitting of excited states (which
up to now has been considered only by a perturbative method \cite{26}) to that of
the ground state.

It has been pointed out that in the low energy region the level shift $%
2\Delta \epsilon _m$ may be obtained by an alternative but equally good
approach called the LSZ method in which only the vacuum instanton (eq. (\ref
{14}) for the problem at hand) plays a role\cite{12,13,15}. The comparison
with the LSZ method is explained in the appendix. The level shift formula
eq. (\ref{35}) is suitable for a quantitative evaluation of the tunneling
amplitude.
We therefore borrow from ref.\cite{5} some data for a ferromagnetic
particle in order to demonstrate the difference in $\lambda$
with respect to the result
of tunneling at the ground state (but we by no means imply
-- for the reasons explained in the introduction -- 
that the transition is a
result of quantum tunneling).

\newpage
\begin{center}
{\bf Table 1}\vspace{0.2in}

Comparison of $\lambda-$values computed for various excited states

\vspace{0.2in}
\begin{tabular}{|c||c|c|c|}
\hline
$m$ & $\lambda $ for $s=3\times 10^3$ & $\lambda $ for $s=2\times 10^3$ & $%
\lambda $ for $s=10^3$ \\ \hline\hline
$2$ & $0.17350\times 10^{-4}$ & $0.37641\times 10^{-4}$ & $1.41226\times
10^{-4}$ \\ \hline
$4$ & $0.24543\times 10^{-4}$ & $0.53458\times 10^{-4}$ & $2.01990\times
10^{-4}$ \\ \hline
$6$ & $0.32066\times 10^{-4}$ & $0.70031\times 10^{-4}$ & $2.65869\times
10^{-4}$ \\ \hline
$8$ & $0.39985\times 10^{-4}$ & $0.87500\times 10^{-4}$ & $3.33361\times
10^{-4}$ \\ \hline
$10$ & $0.48316\times 10^{-4}$ & $1.05899\times 10^{-4}$ & $4.04587\times
10^{-4}$ \\ \hline
$12$ & $0.57064\times 10^{-4}$ & $1.25235\times 10^{-4}$ & $4.79555\times
10^{-4}$ \\ \hline
$20$ & $0.98155\times 10^{-4}$ & $3.15062\times 10^{-4}$ & $20.29119\times
10^{-4}$ \\ \hline
\end{tabular}
\end{center}

The theoretical level shift calculated with the simple model of the WKB
exponential is negligibly small. If the observed frequency reported in ref.%
\cite{5} were considered to be the result of tunneling from the ground
state, $\lambda $ would be $1.01\times 10^{-5}.$ The tunneling effect
enhances significantly at excited states. Table 1 shows how $\lambda $
increases with $m$ (the index of excited states) for tunneling frequency $%
\omega _c=500Hz$ which is the level splitting $4\Delta \epsilon _m$ and 
$s=3\times 10^3,2\times 10^3$ and $10^3$ respectively. For $s=10^3$ to
attain the frequency $\omega _c=500Hz$ the value of $\lambda $ need only be
$2.03\times 10^{-3}$ , which may be a physically acceptable quantity
for a small ferromagnetic particle.

\section{Conclusion}

We have shown that the periodic instanton method as well as the LSZ method
are
useful for the calculation of tunneling effects at excited states of a spin
system at low temperature, the former being valid over the entire region of
energy, the latter at low energies. The results are suitable
for a quantitative
evaluation of the tunneling effect. With the simple model of eq.(\ref{1})
for a ferromagnetic particle we conclude that the tunneling effect at the
level of excited states increases.

\newpage
\begin{center}
{\bf Acknowledgment}
\end{center}

J.--Q.L acknowledges support of the Deutsche Forschungsgemeinschaft and
J.--G.Z.
support of the A. von Humboldt Foundation and the Japan Society
for the Promotion of Science for the award of postdoctoral
fellowships. J.--Q.L., Y.--B.Z. and F.--C.
P. also
acknowledge support of the National Natural Science Foundation of China.

\begin{center}
{\bf Appendix: Comparison with the LSZ result}
\end{center}

From the viewpoint of the LSZ method the transition amplitude between $m$-th
degenerate eigenstates in any two neighboring wells (say those for $n=0,1$)
is viewed as the transition of $m$ bosons induced by the instanton of eq. (\ref
{14}) and is related with the level shift $2\Delta \epsilon _m$ by 
\begin{equation}
A_{f,i}^m=\langle m,\Phi _1|e^{-2\beta \hat{H}}|m,\Phi _0\rangle
=e^{-2\beta\epsilon _m}\sinh 2\beta\Delta \epsilon _m  \eqnum{A.1}  \label{A.1}
\end{equation}
Following ref. \cite{13} the transition amplitude as well as the S-matrix can
be related to the Green's function through the procedure known as the LSZ
reduction technique \cite{14}. To this end we construct the interacting
Euclidean fields $\phi _{\pm }$ in the classically forbidden region with the
help of the instanton configuration of eq.(\ref{14}). Thus we define 
\begin{equation}
\phi _{+}:=\pi -\phi _c,\quad \phi _{-}:=\phi _c  \eqnum{A.2}  \label{A.2}
\end{equation}
such that the interaction fields vanish in their respective asymptotic
regions, i.e. 
\begin{equation}
\lim_{\tau \to \infty }\phi _{+}=0,\lim_{\tau \to -\infty }\phi _{-}=0 
\eqnum{A.3}  \label{A.3}
\end{equation}
The subscripts ``--'' and ``+'' here denote the wells with minima at $\Phi
_0=0$ and $\Phi _1=\pi $ respectively. The Euclidean creation and
annihilation operators $\hat{a}_{\pm }^{\dagger }$ and $\hat{a}_{\pm }$
which create and annihilate an effective boson in wells ``+'' and ``--''
respectively are related to the interaction field operators $\hat{\phi}_{\pm
}$ by 
\begin{eqnarray}
\hat{a}_{\pm }^{\dagger } &=&\sqrt{\frac{2m_0}{\omega _0}}e^{-\omega _0\tau }%
\stackrel{\leftrightarrow }{\frac \partial {\partial \tau }}\hat{\phi}_{\pm
}(\tau )  \nonumber \\
\hat{a}_{\pm } &=&-\sqrt{\frac{2m_0}{\omega _0}}e^{\omega _0\tau }\stackrel{%
\leftrightarrow }{\frac \partial {\partial \tau }}\hat{\phi}_{\pm }(\tau ) 
\eqnum{A.4}  \label{A.4}
\end{eqnarray}
where $m_0 = \frac{1}{2K_1}$, and e.g.
$${\hat a}^{\dagger}_+\stackrel{\tau\rightarrow\infty}{\rightarrow}
-\sqrt{2m_0\omega_0}(1-\lambda )^{-\frac{1}{2}}
$$
The transition amplitude of eq.(\ref{A.1}) can now be written 
\begin{equation}
A_{f,i}^m=S_{f,i}^me^{-2\beta m\omega _0}  \eqnum{A.5}  \label{A.5}
\end{equation}
with S-matrix element 
\begin{equation}
S_{f,i}^m=\lim_{%
{\tau ^i\to -\infty  \atop \tau ^f\to \infty }
}\frac 1{m!}\langle 0|\hat{a}_{+}(\tau _m^f)\ldots \hat{a}_{+}(\tau _1^f)%
\hat{a}_{-}^{\dagger }(\tau _1^i)\ldots \hat{a}_{-}^{\dagger }(\tau
_m^i)|0\rangle   \eqnum{A.6}  \label{A.6}
\end{equation}
The S-matrix element can be evaluated in terms of the Green's function $G$
which arises in its evaluation. Thus 
\begin{eqnarray}
S_{f,i}^m &=&\lim_{%
{\tau ^i\to -\infty  \atop \tau ^f\to \infty }
}\frac 1{m!}\prod_{l=1}^m\left( -\sqrt{\frac{2m_0}{\omega _0}}e^{\omega
_0\tau _l^f}\stackrel{\leftrightarrow }{\frac \partial {\partial \tau _l^f}}%
\right) \left( \sqrt{\frac{2m_0}{\omega _0}}e^{-\omega _0\tau _l^i}\stackrel{%
\leftrightarrow }{\frac \partial {\partial \tau _l^i}}\right) G  \nonumber \\
&=&\frac 1{m!}\prod_{l=1}^m\left( \frac{-2m_0}{\omega _0}\right) e^{\omega
_0(\tau _l^f-\tau _l^i)}\left[ \frac{\partial ^2G}{\partial \tau
_l^f\partial \tau _l^i}+\omega _0\left( \frac{\partial G}{\partial \tau _l^f}%
-\frac{\partial G}{\partial \tau _l^i}-\omega _0G\right) \right]  
\eqnum{A.7}  \label{A.7}
\end{eqnarray}
where the $2m$-point Green's function is defined as usual, i.e. 
\begin{equation}
G=\langle 0|\hat{\phi}_{+}(\tau _m^f)\ldots \hat{\phi}_{+}(\tau _1^f)\hat{%
\phi}_{-}(\tau _1^i)\ldots \hat{\phi}_{-}(\tau _m^i)|0\rangle   \eqnum{A.8}
\label{A.8}
\end{equation}
Following the procedure in refs. \cite{12,13,15} the S-matrix element is thus 
\begin{equation}
S_{f,i}^m=\frac 1{m!}\prod_{l=1}^m\left\{\left( -\frac{2m_0}{\omega _0}\right)
\left[ \frac{d\phi _{+}(\tau _l^f)}{d\tau _l^f}\frac{d\phi _{-}(\tau _l^i)}{%
d\tau _l^i}\right]\right\}A_{f,i}^0  \eqnum{A.9}  \label{A.9}
\end{equation}
where 
\begin{equation}
A_{f,i}^0=2\beta2^2\left\{ \frac{K_1K_2}{(1-\lambda )\pi }\right\} ^{\frac 12%
}\lambda ^{\frac 14}s^{\frac 32}e^{-S_c}e^{-\beta\omega_0}   \eqnum{A.10}  \label{A.10}
\end{equation}
is the amplitude for the transition between degenerate ground states. Using the
instanton solution of eq.(\ref{14}) and taking the large time limit $\tau
_l^i\to -\infty ,\tau _l^f\to \infty $ after performing the imaginary time
derivatives, we finally obtain the transition amplitude by observing that each
pair of vertices in eq. (\ref{A.9}) contributes a factor $-4\omega _0^2$.
Then 
\begin{equation}
A_{f,i}^m=\frac 1{m!}\left(\frac {2^3m_0\omega _0}{1-\lambda}\right)^m
e^{-\omega _0m2\beta}A_{f,i}^0 
\eqnum{A.11}  \label{A.11}
\end{equation}
Comparing eq.(\ref{A.11}) with eq.(\ref{A.1}) we recover the level shift $%
2\Delta \epsilon _m$ of eq.(\ref{35}) for the
low--lying levels. Following refs.\cite{16,17} we here take
$\omega_0 = 2\sqrt{K_1K_2}s$ for reasons of consistency.

\newpage

{\bf Figure Captions}

Fig. 1

The periodic potential and the instanton trajectories:

(a) For one instanton (i.e. vacuum instanton),

(b) For one instanton plus one instanton-anti-instanton pair,

(c) One half of the periodic instanton.

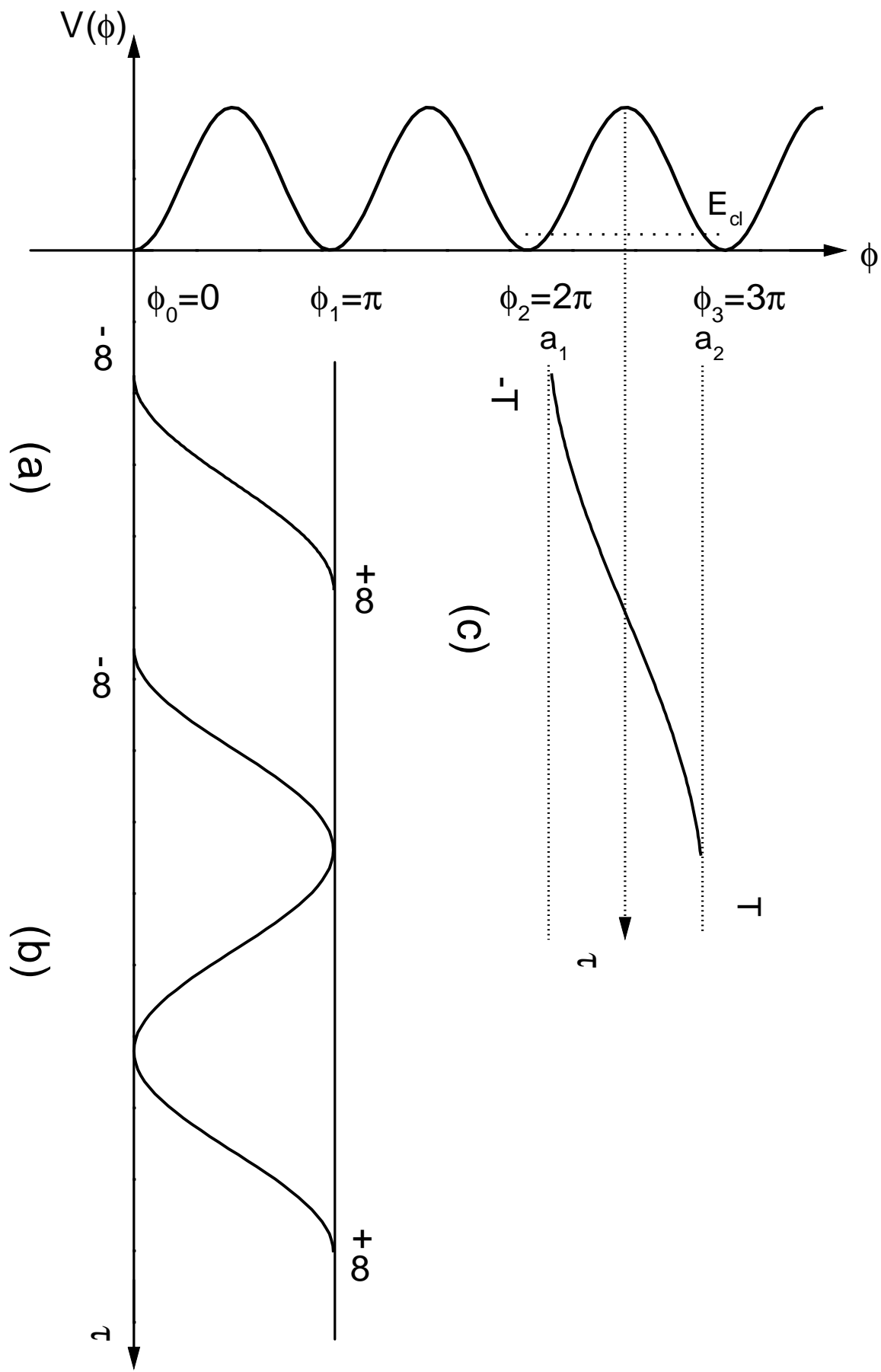
\end{document}